\documentclass[fleqn,usenatbib]{mnras}
\usepackage{newtxtext,newtxmath}
\usepackage[T1]{fontenc}
\DeclareRobustCommand{\VAN}[3]{#2}
\let\VANthebibliography\thebibliography
\def\thebibliography{\DeclareRobustCommand{\VAN}[3]{##3}\VANthebibliography}

\usepackage{graphicx}	
\usepackage{amsmath}	

\title[Mergers in NGC\,7135]{Old and New Major Mergers in the SOSIMPLE galaxy, NGC\,7135}

\author[T. Davison et al.]{
Thomas A. Davison,$^{1,2}$\thanks{E-mail: tdavison@eso.org}
Harald Kuntschner,$^{1}$
Bernd Husemann,$^{3}$
Mark A. Norris,$^{2}$
\newauthor
\,Julianne J. Dalcanton,$^{4}$
Alessandra De Rosa,$^{5}$
Pierre-Alain Duc,$^{6}$
Stefano Bianchi,$^{7}$
\newauthor
\,Pedro R. Capelo,$^{8}$
Cristian Vignali$^{9,10}$
\\
$^{1}$European Southern Observatory, Karl-Schwarzschild-Strasse 2, D-87548 Garching bei Muenchen, Germany\\
$^{2}$Jeremiah Horrocks Institute, University of Central Lancashire, Preston PR1 2HE, UK\\
$^{3}$Max-Planck-Institut für Astronomie, Königstuhl 17, D-69117 Heidelberg, Germany\\
$^{4}$Department of Astronomy, University of Washington, Box 351580, Seattle, WA 98195, USA\\
$^{5}$INAF - Istituto di Astrofisica e Planetologie Spaziali, Via Fosso del Cavaliere, 00133 Rome, Italy\\
$^{6}$Universit\'{e} de Strasbourg, CNRS, Observatoire astronomique de Strasbourg (ObAS), UMR 7550, 67000 Strasbourg, France\\
$^{7}$Dipartimento di Matematica e Fisica, Università degli Studi Roma Tre, via della Vasca Navale 84, I-00146 Roma, Italy\\
$^{8}$Center for Theoretical Astrophysics and Cosmology, Institute for
Computational Science, University of Zurich,\\ Winterthurerstrasse 190,
CH-8057 Z{\"u}rich, Switzerland\\
$^{9}$Dipartimento di Fisica e Astronomia, Alma Mater Studiorum, Universit\`a degli Studi di Bologna, Via Gobetti 93/2, I-40129 Bologna, Italy\\
$^{10}$INAF - Osservatorio di Astrofisica e Scienza dello Spazio di Bologna, Via Gobetti 93/3, I-40129 Bologna, Italy
}
\date{Accepted XXX. Received YYY; in original form ZZZ}
\pubyear{2020}

\begin{document}
\label{firstpage}
\pagerange{\pageref{firstpage}--\pageref{lastpage}}
\maketitle

\begin{abstract}
The simultaneous advancement of high resolution integral field unit spectroscopy and robust full-spectral fitting codes now make it possible to examine spatially-resolved kinematic, chemical composition, and star-formation history from nearby galaxies. We take new MUSE data from the Snapshot Optical Spectroscopic Imaging of Mergers and Pairs for Legacy Exploration (SOSIMPLE) survey to examine NGC\,7135. With counter-rotation of gas, disrupted kinematics and asymmetric chemical distribution, NGC\,7135 is consistent with an ongoing merger. Though well hidden by the current merger, we are able to distinguish stars originating from an older merger, occurring 6-10\,Gyr ago. We further find a gradient in ex-situ material with galactocentric radius, with the accreted fraction rising from 0\% in the galaxy centre, to $\sim$7\% within 0.6 effective radii.
\end{abstract}


\section{Introduction}
Galaxy merger research has shown how fundamental merging is to galaxy evolution, with historical merger rates generally increasing with galaxy mass \citep{bundy2009greater, schawinski2010role, l2012mass, pillepich2018first}.  Distant galaxies (z$\approx$2) are often quoted as being a factor of 2-5 times smaller than those found locally \citep{daddi2005passively,van2008confirmation, saracco2009population}. As such it is widely assumed that a large amount of mass-assembly after z$\approx$2 is a result of hierarchical growth through galaxy mergers and accretion which has been widely corroborated from galaxy evolution models. Not only does merger history impact on almost all other aspects of galaxy evolution, but many galaxies have experienced large mergers throughout their history with around 50\% of galaxies experiencing a major merger \citep{maller2006galaxy}, and essentially all surviving galaxies experiencing minor mergers, with frequency increasing with merger mass-ratio \citep{lotz2011major}. The exception for these cases are some rare pristine galaxy types ($\lesssim$ 0.1\% of galaxies according to \citealt{quilis2013expected}) which have likely experienced no outside interaction or accretion events \citep{trujillo2013ngc}.

Modelling is an excellent way to delve into the mechanics and subsequent effects of galaxy mergers. Using simulations, the ex-situ mass fraction of accreted galaxies has been explored in depth \citep{pillepich2015building, qu2017chronicle, davison2020eagle}. This is useful for defining expected current merger rates to be compared to observationally. A challenging aspect of observational astronomy is demonstrating the merger history of observed nearby galaxies to verify these models, particularly if potential mergers occurred several Gyr ago.

Integral Field Spectroscopy has proven particularly useful in exploring galaxy kinematics and populations. Integral Field Units (IFU's) have provided spatially resolved maps of galaxies which can be used to diagnose population differences and kinematic effects as a result of mergers. This has been shown to be effective in numerous observational cases \citep[see e.g.][]{guerou2016, faifer2017, Ge2019}

The impact of mergers and merger history on galaxy evolution is an important aspect to understand. For one thing, mergers are known to drive gas towards the galaxy centre \citep{mihos1995gasdynamics}, causing AGN activity and black hole growth, which in turn can shut down or suppress star formation in the galaxy at large \citep{cales2015post, choi2015impact}. On the other hand, mergers can cause sudden and significant bursts of star formation due to the disruption of previously unperturbed gas kinematics \citep{di2008frequency, ellison2013galaxy, moreno2015mapping, capelo2015growth}. Disruption in the gas kinematics of galaxies can leave key fingerprints in identification of merger events. One of the most readily identifiable features of a recent or ongoing merger is counter rotating components, with up to 40\% of S0 galaxies displaying signatures of counter-rotation \citep{rubin1994multi, davis2011atlas3d, coccato2015properties, bassett2017formation}. Galaxy-galaxy mergers of the right combination can change the very morphological type of a galaxy. As such, mergers hold the power to define entire galaxy futures.

The S01-pec galaxy NGC\,7135 (AM 2146–350, IC 5136) in the constellation of Piscis Austrinus is a merger remnant galaxy \citep{Keel1985} that is likely en route to forming an S0 galaxy. It currently displays several immediately striking visual features including an extended tail, shell features, and curved structure (Figure \ref{phot}) based on photometry from the Carnegie-Irvine Galaxy Survey \citep{ho2011carnegie}.

NGC\,7135 was first described as having `a curious jet and shell' in \cite{malin1983catalog} with the `jet' later shown to be a tail in \cite{2003MNRAS.343..819R}. The shell structures of the galaxy were found to be particularly clear in UV \citep{rampazzo2007, marino2011nearby}, with FUV gas structure further linked to an accretion event that also likely formed the shells. \cite{ueda2014cold} found CO emitting gas that was unassociated with the nucleus, along with 3 mm continuum associated with the nucleus. Despite speculation, NGC\,7135 was determined to have no active nucleus as shown in \cite{zaw2009galaxies} through optical spectra analysis. Analysis in \cite{1985keel} identifies NGC\,7135 as a merger galaxy, and in \cite{2003MNRAS.343..819R} NGC\,7135 is shown to possess an elongated, asymmetric gas structure relative to the stellar material. 

The local environment of NGC\,7135 is described by \cite{samir2016fundamental} as being `low density', with the classification of `low density' \citep{annibali2010nearby} a result of the richness parameter $\rho_{xyz}$=0.32 gal Mpc$^{-3}$ \citep{tully1988nearby}. Early type galaxies in low density environments are known to possess on average younger populations ($\sim$\,2Gyr younger) than similar galaxies in higher density environments \citep{thomas2003stellar}, a likely result of more recent mergers and star formation.

In this work we present new observations of the galaxy NGC\,7135, recently obtained with MUSE. We aim to show that NGC\,7135 is currently undergoing a major merger, with a history of older mergers underlying in the galaxy populations. The paper is presented as follows: In Section 2 we describe the motivation behind the observations, as well as the data reduction and limitations. In Section 3 we describe our methodology, including the use of regularisation during spectral fitting. In Section 4 we present the resultant maps of stellar populations and kinematics, as well as gas properties similarly derived, including rotation differences between the two components. In Section 5 we discuss the implications of the results and finally in Section 6 we provide a summary and concluding remarks.

\section{Observations and data reduction}
We observed NGC\,7135 with the Multi Unit Spectroscopic Explorer \citep[MUSE,][]{bacon2010MUSE,bacon2014MUSE} at the Very Large Telescope (VLT) as part of the Snapshot Optical Spectroscopic Imaging of Mergers and Pairs for Legacy Exploration (SOSIMPLE) survey (Program ID: 0103.A-0637(A), PI: B.~Husemann). The aim of the SOSIMPLE survey is to provide complementary IFU observations for an ongoing Hubble filler gap snapshot imaging program (Program ID: 15446, PI: J.~Dalcanton). HST imaging of NGC\,7135 is not yet taken due to the filler nature of the HST program, thus these MUSE observations act as a first look at the data, to which HST data can be compared to at a later date. Combining IFU spectroscopy with a large set of high-quality ancillary data will hopefully provide observational and theoretical insights into the evolution of merging systems.

The MUSE observations were conducted on 6 July 2019 during dark sky conditions and split into 3$\times$560\,s dithered pointings along with a 300\,s dedicated blank sky field exposure for background subtraction of this extended galaxy. Rotations of 90\degr\ were applied between exposures covering approximately 3.4 arcmin$^2$ as shown in Fig \ref{phot}. The seeing during the observations maintained at $\sim$1\arcsec\ , and the sky was covered with thin clouds during strong wind conditions from North-West direction.  

The data were reduced with the standard ESO pipeline \citep{weilbacher2020pipeline} which performs detector calibrations, flat-fielding, wavelength calibration, flux calibration as well as sky subtraction, exposure alignment, and cube reconstruction of the combined exposures. We performed an additional correction for residual sky lines using a simple PCA algorithm. The MUSE pixel scale is 0.2 arcsec pixel$^{-1}$, with a mean spectral resolution of $\sim$2.5\AA\ though this can vary across the wavelength range (see figure 5 of \citealt{husser2016muse}). The resulting mean Signal-to-Noise (SN) ratio of the spaxels in the MUSE image within a wavelength range of 4759--6849\,\AA\ (limited from 4759--9300\,\AA ) is 9.5, with a maximum spaxel SN of 131.

\section{Methodology}\label{method}
Spaxels were Voronoi binned to a minimum SN of 50 per \AA, thereby poor signal regions were made available for analysis, whilst higher SN spaxels remained unbinned. This optimally allowed for spatial investigation of spectral properties, without losing valuable high resolution data at high SN locations.

\begin{figure}
    \includegraphics[width=\linewidth]{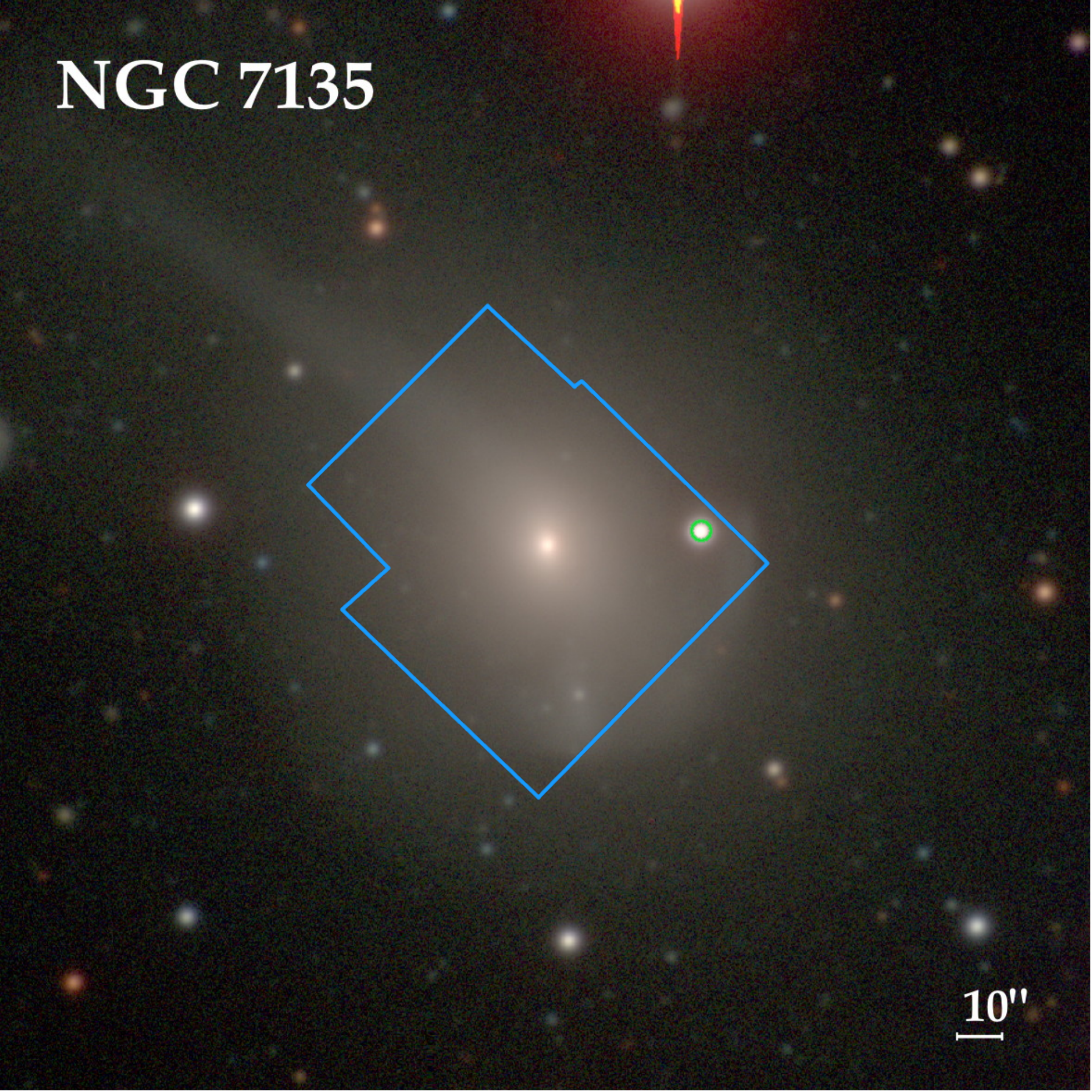}
    \caption{A colour image of NGC\,7135 showing the MUSE cube footprint. Photometry of NGC\,7135 is from the Carnegie-Irvine Galaxy Survey \citep{ho2011carnegie}. The blue border shows the boundaries of the reduced MUSE IFU data used in this study. 
    A green circle traces an area containing a bright foreground star that was entirely excluded from the analysis.}
    \label{phot}
\end{figure}

The wavelength was restricted to 4759 - 6849\,\AA\, for all spaxels to ensure the strongest Balmer lines were included, and to exclude noisier sky-dominated regions at redder wavelengths. All spectra of spaxels within a bin were summed into a single spectra representing the area covered by the bin. An area containing a foreground star was masked from analysis in the West of the image (see Figure \ref{phot}). 

To analyse the spectra from the binned NGC\,7135 data we utilised the Penalized PiXel-Fitting (pPXF) method, described in \cite{cappellari2004intro} and upgraded in \cite{cappellari2017upgrade}. With this method, single-age single-metallicity stellar population (SSP) models are fit to spectra to build a map of stellar populations across age and metallicity space. By identifying the combination of SSP models that approximate a given spectrum, the estimated constituent populations are extracted, as well as velocity and dispersion. Stellar models are weighted as per the estimated fraction of the population present in the galaxy. As a result, output weights of stellar models indicate the fractions of specific stellar populations present in the spectrum. The output model of combined spectra is made more physical by the use of template regularisation (see e.g. section 3.5 of \citealt{cappellari2017upgrade}), the methodology of which is explained in detail below. Standard pPXF cleaning algorithms were included to mask emission lines where necessary.

A total of 552 MILES SSP models \citep{vazdekis2010evolutionary} were used to fit to galaxy spectra. These models were of Kroupa revised initial mass function (log slope of 1.3, M$_{max}$=100M$_{\odot}$) using BaSTI isochrones, with a metallicity range of -2.27 to +0.4 [M/H] in 12 non-linear steps, and an age range of 0.1 to 14.0\,Gyr in 46 non-linear steps \cite[][]{kroupa2001variation, cassisi2005basti,pietrinferni2006large,falcon2011updated,vazdekis2012miuscat}.

Application of regularisation allows smoothing over stellar model weights to reproduce a population map consistent with physical results. The weighted templates that have been combined to produce a target spectrum will often be unphysically localised to only the strongest of possible solutions, with many other valid solutions being overlooked, despite their physicality. To produce more representative distributions, regularisation seeks to smooth the solutions to a physical state. The challenge is to smooth the template weights to a solution that most accurately represents observed conditions, whilst not overlooking genuine fluctuations and details present in the model-fit. The regularisation parameter controls the strength of the smoothing and is deduced through a robust iterative approach for each spectrum individually. The regularisation parameter is derived such that it corresponds to the maximum value consistent with observations. Thus the derived star formation history will be the smoothest that is consistent with the observations. This has been shown in literature to be an accurate and useful method of galaxy population extraction \cite[see e.g.][]{comeron2015, norris2015extended, guerou2016, faifer2017, Ge2019, boecker2020recovering}.

In this work an iterative routine is applied to extract the optimal regularisation parameter. For the best possible fit, the $\chi^2$ of the solution is expected to be approximately equal to the number of available voxels in the spectrum, $N$ (i.e. the number of voxels available after any masking). To obtain this optimal solution, the $\chi^2$ must be increased from the unregularised $\chi^2$ (referred to as $\chi^2_0$) by $\sqrt{2N}$. 

After rescaling noise from the unregularised solution such that $\frac{\chi^2}{N}$ = 1, we make a number of primary guesses at the regularisation parameter. We find the $\Delta \chi^2$ of these initial guesses and fit a function to the input regularisation guesses and output $\Delta \chi^2$ values. By doing so we can precisely find the optimal regularisation parameter such that $\chi^2 = \chi^2_0+\sqrt{2N}$. This action is performed for every bin, resulting in optimal solutions across the entire image map.

\section{Results}
We separate the analysis of NGC\,7135 into three components; the stellar component analysis, encompassing the stellar kinematics; the gaseous component analysis, encompassing gas kinematics, emission lines and star formation aspects; and the population analysis, examining the various stellar populations and the resulting implications for the assembly history of NGC\,7135.

To examine the stellar component we utilise Voronoi binning as described in Section \ref{method}. From this we are able to examine the stellar rotation and bulk velocities, as well as mean age and metallicities spatially across the galaxy (Fig \ref{stellar_properties}). To investigate details related to the gaseous component we use regular binning to view the gas velocities and rotation, as well as the line strengths of H$\alpha$ and H$\beta$ (Fig \ref{gas_properties}). Though we see reasonable amounts of H$\alpha$ emission, there is scant evidence for significant ongoing star formation. This is explained in detail in Section \ref{gas_props}. Finally, in Section \ref{stell_pops_text} we further analyse age and metallicity distributions for sampled regions across the galaxy to diagnose assembly history and current merger status, then go on to examine underlying metal poor populations in Section \ref{acc_pop}.

\subsection{Stellar Properties}
\begin{figure*}
    \includegraphics[width=\linewidth]{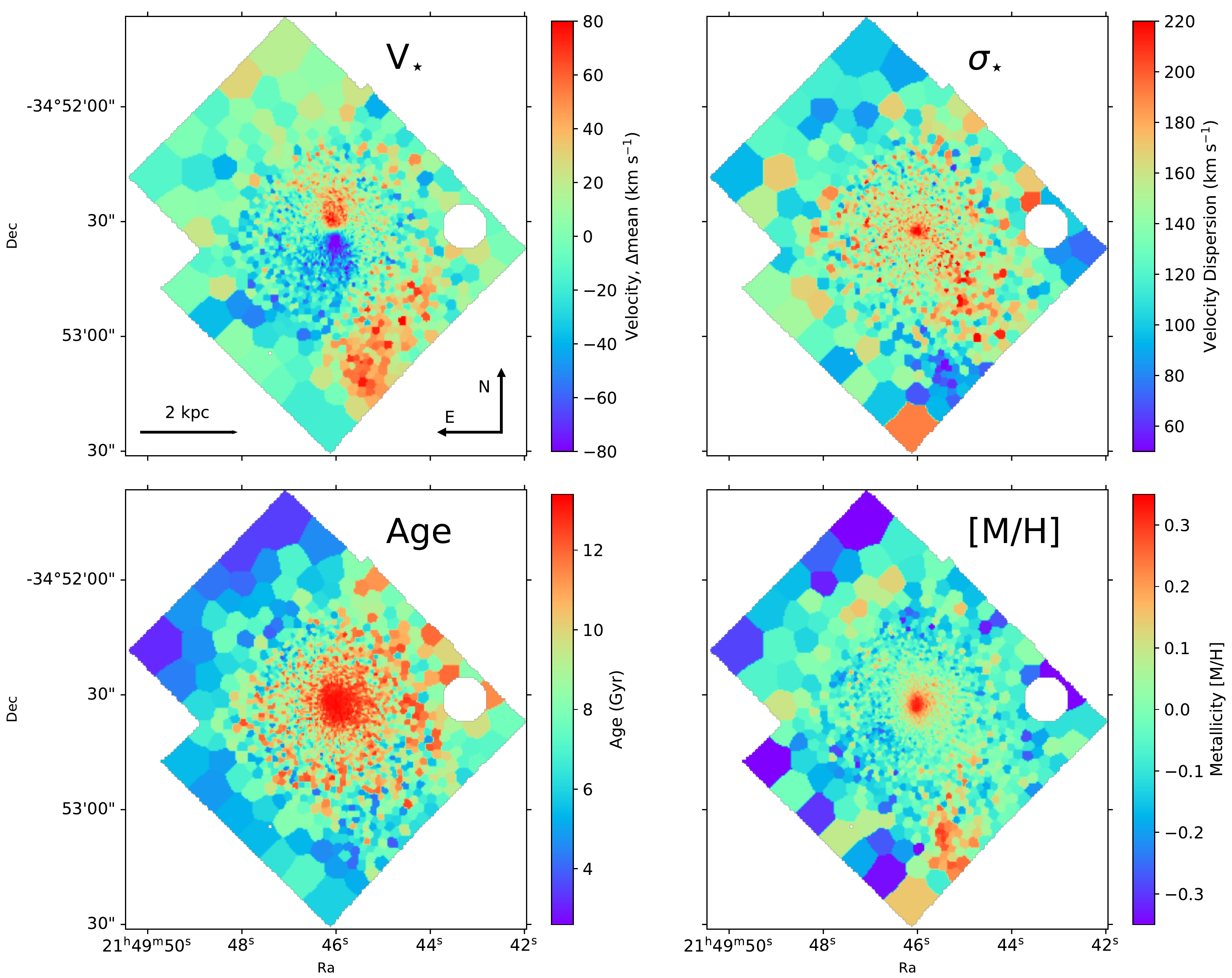}
    \caption{Voronoi map of NGC\,7135 showing 4 different stellar kinematic or mass-weighted population properties. The top left panel shows the mean velocity in km/s for each bin. The top right panel shows mean velocity dispersion within bins in km/s. The lower left panel shows the mean age of populations within the bin in Gyr. Finally the lower right panel shows mean metallicity within each bin. North is to the top of the image, and East is to the left. The stars show clear rotation in the centre. Velocity dispersion, age and metallicity all increase towards the galaxy centre. Distinct kinematics and metallicity south of the centre highlight a distinct component.}
    \label{stellar_properties}
\end{figure*}

Application of the pPXF method to the NGC\,7135 data cube provides mean kinematic properties which are extracted from each bin. Demonstrations of this for velocity and velocity dispersion of the galaxy are found in the top panels of Figure \ref{stellar_properties}. Application of regularisation and mass-to-light ratios produce maps of the constituent stellar populations within each bin of the galaxy. From these bins we can derive mean mass-weighted stellar age and metallicity values, as demonstrated in the lower panels of Figure \ref{stellar_properties}.

The stellar kinematic, age, and metallicity maps of NGC\,7135 reveal much about the galaxy. Stellar rotation is immediately visible. This is of key interest when comparing to gas which rotates counter to the direction of stellar rotation. This is explored in detail in Section \ref{gas_props}. One prominent kinematic feature, perhaps most clearly seen in the velocity map (top left panel) of Figure \ref{stellar_properties}, is an arc of incongruous material at higher than average velocity, stretching from the South West of the Figure to the West. The Southern end of this arc is matched in the metallicity map (lower right panel, Figure \ref{stellar_properties}) by a higher metallicity region, which is also distinct in velocity and velocity dispersion. Upon inspection, this is revealed to be an infalling galaxy currently merging onto NGC\,7135. This can be clearly seen in photometry shown in Figure \ref{var_regions}, and even more compelling evidence comes from population analysis below. 

\subsection{Gas Properties}\label{gas_props}
\begin{figure*}
    \includegraphics[width=\linewidth]{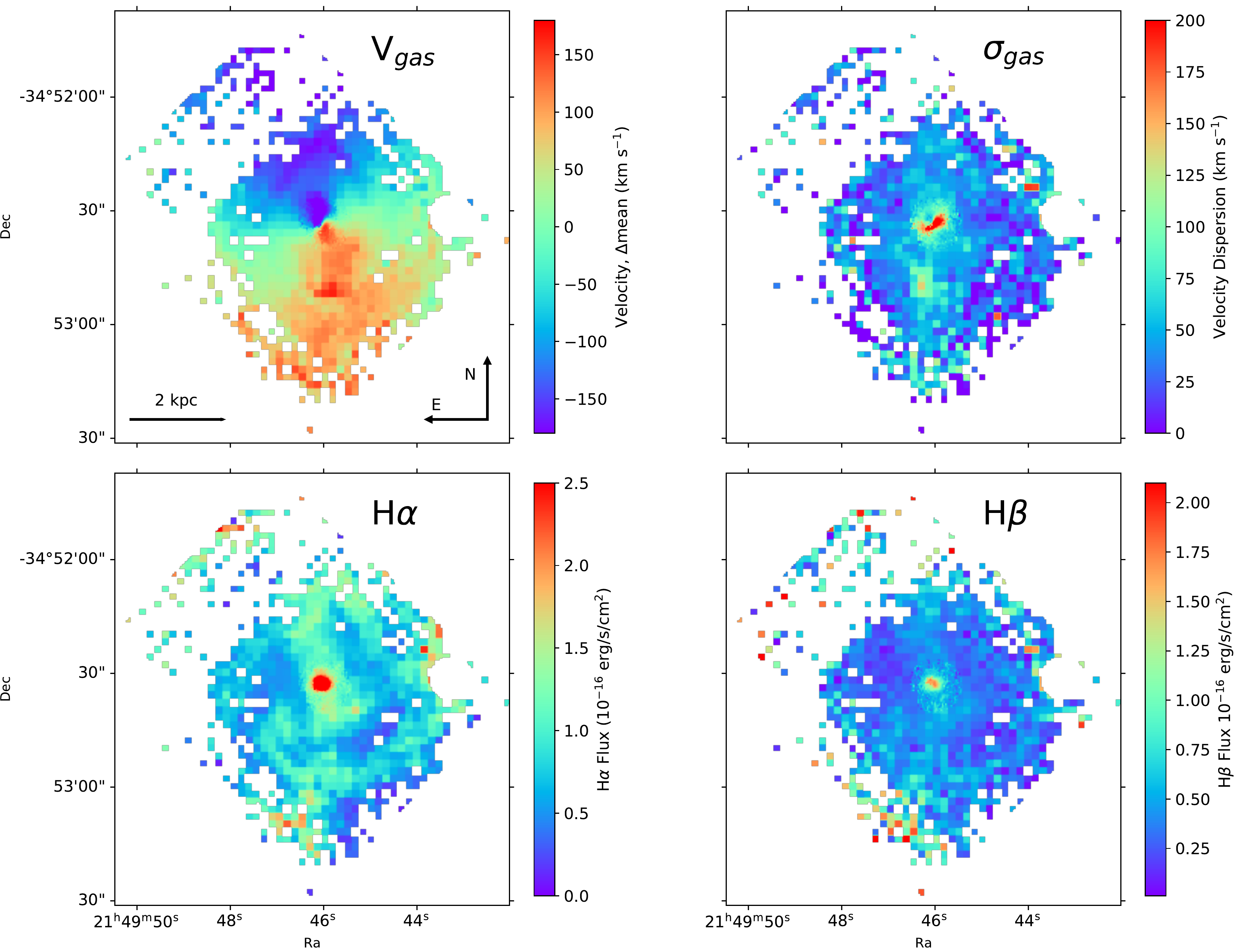}
    \caption{Regularly binned map of NGC\,7135 showing 4 different gas kinematic and strength properties. The top left panel shows the mean velocity of gas in km/s for each bin. The top right panel shows mean velocity dispersion of gas within bins in km/s. The lower left panel shows the H$\alpha$ flux throughout NGC\,7135. The scale has been limited from the true maximum to better display regions of intermediate strength. This limits the core from a true strength of at most 36.2$\times$10$^{-16}$erg/s/cm$^2$ (limited to 2.5$\times$10$^{-16}$erg/s/cm$^2$). The lower right panel shows H$\beta$ flux throughout NGC\,7135. The scale has been limited from the true maximum to better display regions of intermediate strength. This limits the core from a true strength of at most 5$\times$10$^{-16}$erg/s/cm$^2$ (limited to 2.1$\times$10$^{-16}$erg/s/cm$^2$). The gas velocity shows counter rotation compared to the stellar component, and on a slightly different axis, suggesting a merger origin. }
    \label{gas_properties}
\end{figure*}
\begin{figure*}
    \includegraphics[width=\linewidth]{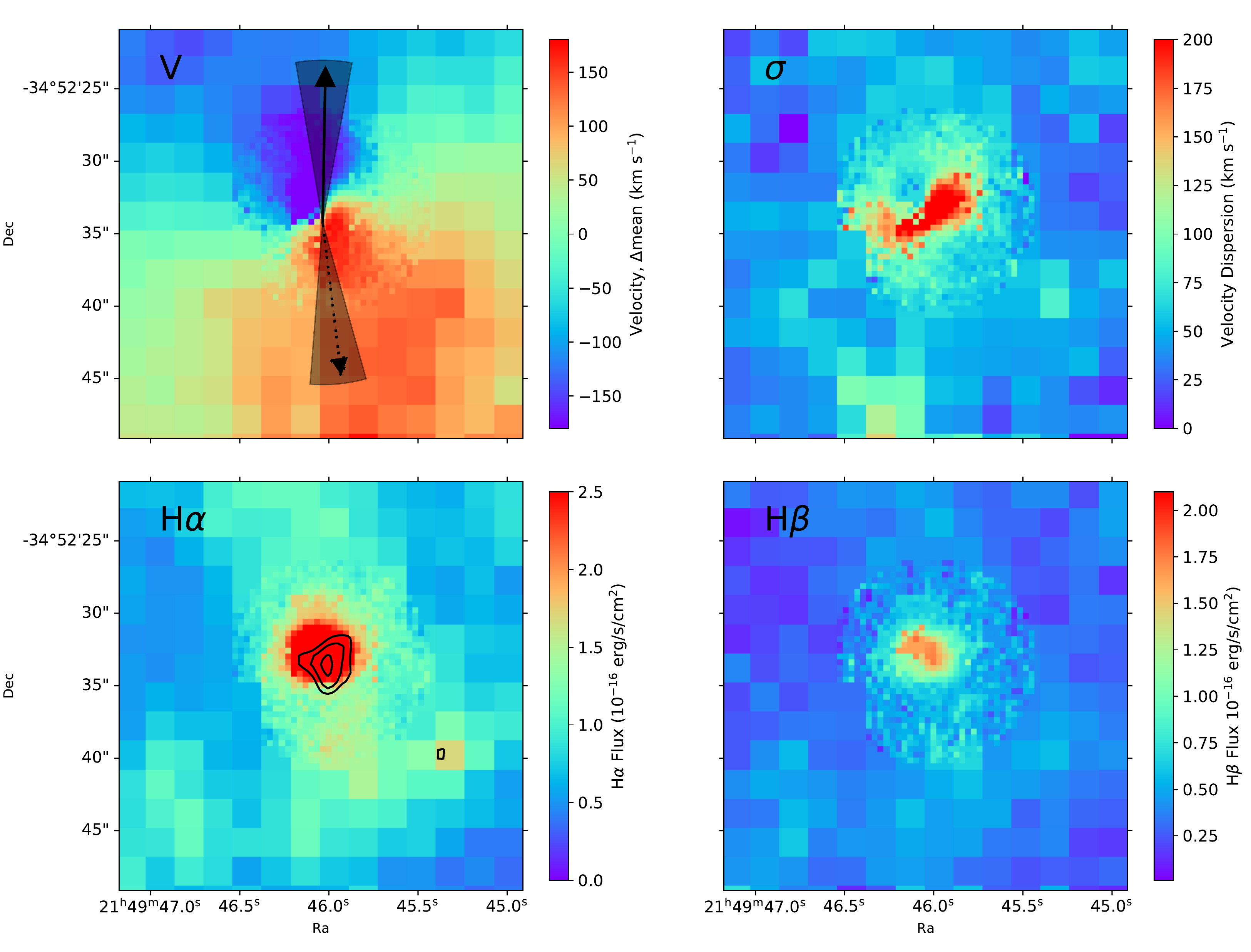}
    \caption{Regularly binned and zoomed in map of NGC\,7135 showing 4 different gas kinematic and strength properties. The top left panel shows the mean velocity of gas in km/s for each bin. The top right panel shows mean velocity dispersion of gas within bins in km/s. The lower left shows the H$\alpha$ flux throughout NGC\,7135. The scale has been limited from the true maximum to better display regions of intermediate strength. This limits the strongest emission near the core from a true strength of at most 36.2$\times$10$^{-16}$erg/s/cm$^2$ (limited to 2.5$\times$10$^{-16}$erg/s/cm$^2$). The lower right panel shows H$\beta$ flux throughout NGC\,7135. The scale here has also been limited. This limits the strongest emission from a true strength of at most 5$\times$10$^{-16}$erg/s/cm$^2$ (limited to 2.1$\times$10$^{-16}$erg/s/cm$^2$). In the upper left panel, arrows show the average positive rotation direction. The solid arrow indicates the average stellar component positive rotation whilst the dotted arrow shows the average gas positive rotation direction. Shaded regions show the standard deviation of vectors for both components for bins of 0.1 effective radii. In the lower left panel, contours show integrated CO(J=1–0) emission detected in ALMA observations \citep{ueda2014cold}. Contours show the 0.8, 1.0 and 1.2 Jy km s$^{-1}$ levels. There is pervasive H$\alpha$ emission with a high luminosity and high velocity dispersion component in the centre, though there is little evidence of star formation.}
    \label{gas_properties_zoom}
\end{figure*}
To explore gas kinematics and distribution in NGC\,7135, regular binning was employed to avoid biases caused by the stellar light controlling Voronoi binning. Large square bins containing 64 pixels were selected across the face of the data cube, and spectra within a given bin were summed and analysed with ppxf as described in Section \ref{method}. Following this, those bins with signal-to-noise that exceeded the minimum detection threshold were re-binned to a higher resolution. This adaptive `zoom' binning gave high resolution in areas of strong H$\alpha$ emission. The zoom resolution was limited to central regions of the galaxy, where the finest detail was required.

NGC\,7135 displays localised areas of strong Balmer emission, shown in Figure \ref{gas_properties} with a cropped version showing the galaxy centre in Figure \ref{gas_properties_zoom}. As seen from all panels, the gas is asymmetric in distribution as well as in kinematics. The rotation of the gas highlights the decoupled nature of the stellar material in the core. 

Gas is counter-rotating to the stellar component, strongly indicating a disrupted system. A slight deviation to the coherent gas movement is seen in the galaxy centre, giving an `S' shaped gas rotation profile. Counter rotation has long been associated with galaxy mergers \citep[see e.g.][]{bertola1988counter}. Total decoupling of gas rotation from stellar components as a result of prograde-prograde merger shocks has been shown in simulation in \cite{capelo2016shocks}, and a similar event appears to be in play here, wherein a major merger has resulted in a counter rotation of the gas component. Plausibly this is the result of a previous merger providing counter rotation from a prograde-prograde merger, this is expanded further in section \ref{stell_pops_text}. Alternatively, counter rotation could have arisen as a result of a first pass of the currently infalling galaxy.

Velocity vectorisation of the gas and stars allows us to measure the gas and stellar rotation misalignment. The rotation consensus in the gas is fairly standard, with the gas rotating around the centre. In the stellar component however, matters are complicated by the velocity of the in-falling galaxy, which shifts the positive rotation vector compared to the core. If we consider only the core, the misalignment of gas and stars is 176$^{\circ}$, whereas when the entire cube is considered, the misalignment is 139$^{\circ}$. This is entirely within the realm of expected values for an interacting galaxy \citep[see e.g.][]{barrera2015tracing, bryant2019sami}. This is shown in Figure \ref{gas_properties_zoom} as solid and dashed arrows for the directions of mean positive stellar and gas rotation respectively, with associated errors shown as shaded regions.

Regions of H$\alpha$ emission can be seen in the southern areas of the lower left panel of Figure \ref{gas_properties}. This forms a large arc with patches exhibiting particularly strong emission. These are seemingly matched by arcs in the north in an asymmetrical manner.

Considering the gas asymmetry and the increase in both gas velocity and velocity dispersion, a large amount of gas can be attributed to material stripped from the outskirts of the infalling galaxy and which is currently in the process of accreting onto the host galaxy. This is seen in the largest area of gas velocity dispersion occurring outside the core, located in a tight region south of the galaxy core. This region indicates a quantity of gas that is not associated with the cohort gas of NGC\,7135, as it displays a region where infalling gas is interacting with the galaxy interstellar medium. This area of higher than expected dispersion is in the plane of the galaxy gas rotation, again evidence that gas is infalling, creating high velocity dispersion at the position where in-situ gas meets ex-situ gas.

\begin{figure}
    \includegraphics[width=\linewidth]{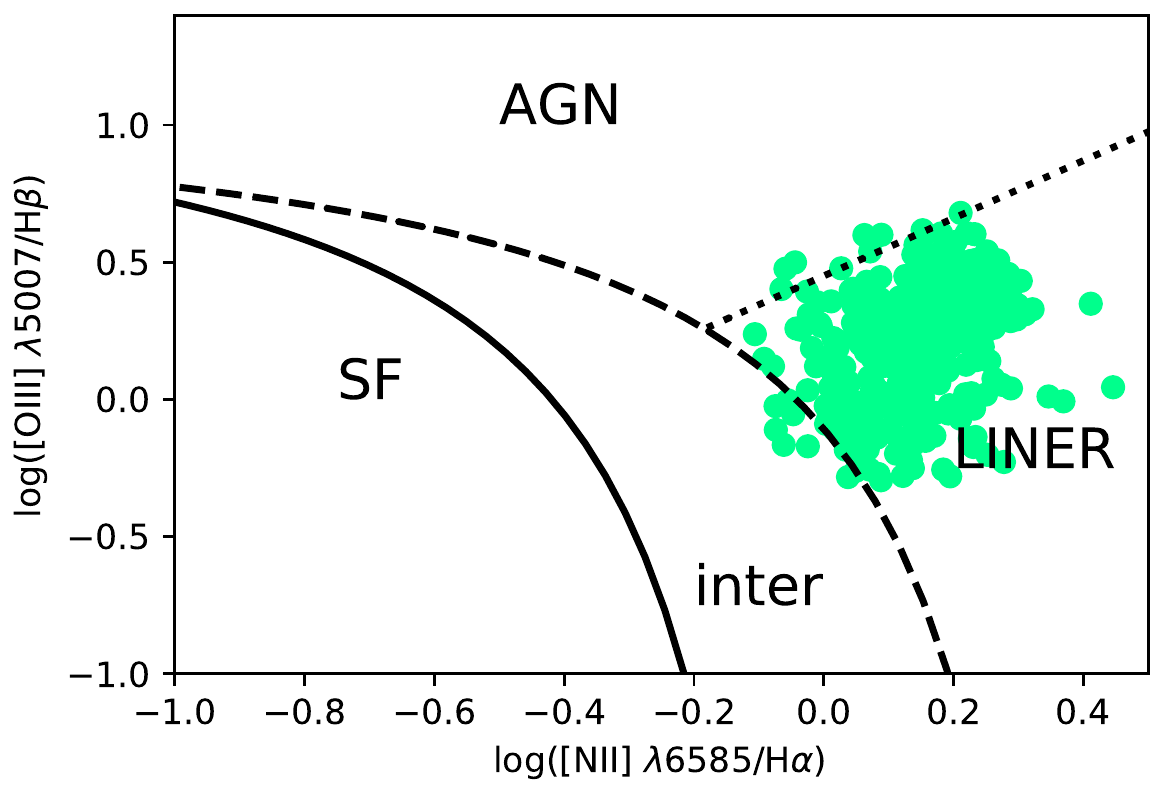}
    \caption{An emission line diagnostic diagram \citep{baldwin1981classification} divided into various sources. Each bin is shown as a point according to its emission ratios of [NII]/H$\alpha$ and [OIII]/H$\beta$ allowing for the identification of regions of star formation, AGN emission or Low-ionization nuclear emission-line region (LINER) emission. Detailed description of the line equations can be found in \protect\cite{park2013relationship}. NGC\,7135 shows no bins where current star formation is clear in the emission. Slight overlaps outside the LINER emission bin are unlikely to be genuine, but rather likely arise because of noise and intrinsic variations. The galaxy emission is overwhelmingly LINER type.} 
    \label{bpt}
\end{figure}

A strong presence of H$\alpha$ in concentrated regions is consistent with the picture of NGC\,7135 as a galaxy that has perhaps recently undergone star formation as suggested in \cite{rampazzo2007}, though at low levels. Despite this, there is little to no evidence of strong ongoing star formation. This can be seen in the emission line diagnostic diagram in Figure \ref{bpt}. Almost all the sources of emission are associated with low-ionization nuclear emission-line regions (LINERs). Though a handful of active galactic nuclei (AGN) sources can be seen, they largely lie in the outer noisier regions of the data-cube, which makes the presence of true AGN sources doubtful, as shown in \cite{zaw2009galaxies}. This strong bias towards LINER emission is typical of merging systems with shock driven LINER emission \citep{monreal2010vlt, rich2011galaxy}.

ALMA data \citep{ueda2014cold} showing the $^{12}$CO(J=1–0) emission is overlaid in the lower left panel of Figure \ref{gas_properties_zoom}. The ALMA observations reveal a significant peak in CO emission offset from the galaxy core with an integrated molecular gas mass of $M_{\mathrm{H2}}=(5.4\pm1.4)\times10^7M_{\sun}$ adopting an $\alpha_\mathrm{CO}=4.8M_{\sun}\,\mathrm{pc}^{-2}(\mathrm{K\,km\,s}^{-1})^{-1}$ \citep{solomon1991co}. This cold gas mass would correspond to an expected SFR of only $\sim0.025M_{\sun}\,\mathrm{yr}^{-1}$ if a normal depletion time of 2\,Gyr for galaxies is assumed \citep{bigiel2011constant, leroy2013molecular}. Although there is no similarly distinct ionised gas structure observed with MUSE, there is plenty of ionized gas which may partially originate from star formation despite the LINER-like classification. The extinction-corrected H$\alpha$ flux within the central r=1$\arcsec$ is $(4\pm0.4)\times10^{-13}\mathrm{erg}\,\mathrm{s}^{-1}\,\mathrm{cm}^{-2}$ which would correspond to $\mathrm{SFR}=0.5\pm0.05M_{\sun}\,\mathrm{yr}^{-1}$ following \citet{kennicutt1998global}. So only 5\% of the central H$\alpha$ would need to be hidden among LINER-like classified ionised gas to be in agreement with ongoing star formation. Such a low fraction of star formation would not alter the line diagnostics significantly and would remain hidden. Hence, we cannot rule out ongoing star formation based on the central cold gas mass  observed by \cite{ueda2014cold}. Given the highly disturbed kinematics, the possibility that dynamical suppression of star formation is preventing cold gas collapse cannot be tested by our observations.

\subsection{Stellar Population Mapping}\label{stell_pops_text}
Populations of a galaxy evolve in metallicity over time, gradually enriching with age. The exact quantities and rates of this enrichment are well known \citep{carraro1994galactic,Layden_2000,pont2004}, with the rate of enrichment directly tied to galaxy mass resulting in the mass-metallicity relation. Thus, we can quickly establish whether a galaxy has followed the standard enrichment of its population as would be expected from an isolated galaxy.

In reality, galaxies are more often than not experiencing regular disturbances in the form of mergers, fly-bys and intracluster medium interaction such as ram-pressure stripping \citep{lotz2011major, sinha2012first, ebeling2014jellyfish, ventou2017muse}. One effect of this is the variation of the age-metallicity relation of a galaxy from the modelled form. This is most strikingly clear when a galaxy accretes material from a lower mass galaxy \citep{spolaor2009mass, leaman2013bifurcated}. Due to the lower metal enrichment rate of lower mass galaxies than that of larger mass galaxies, one finds that in general a smaller mass galaxy will exhibit far lower values of metallicity at late ages. Because of the ability for full spectral fitting methods to identify populations based on age and metallicity models, one would see these two populations as distinct and separate areas on an age-metallicity diagram. This is dependent on the difference in mass of the mergers however, as if two galaxies of similar mass were to merge, the separation of populations on the age-metallicity diagram would be too little to distinguish at the current resolutions of full-spectral fitting methods. Using these principles we can estimate which of the populations present are those which have accreted onto the host galaxy, and are therefore ex-situ in origin.

\begin{figure*}
 \includegraphics[width=\linewidth]{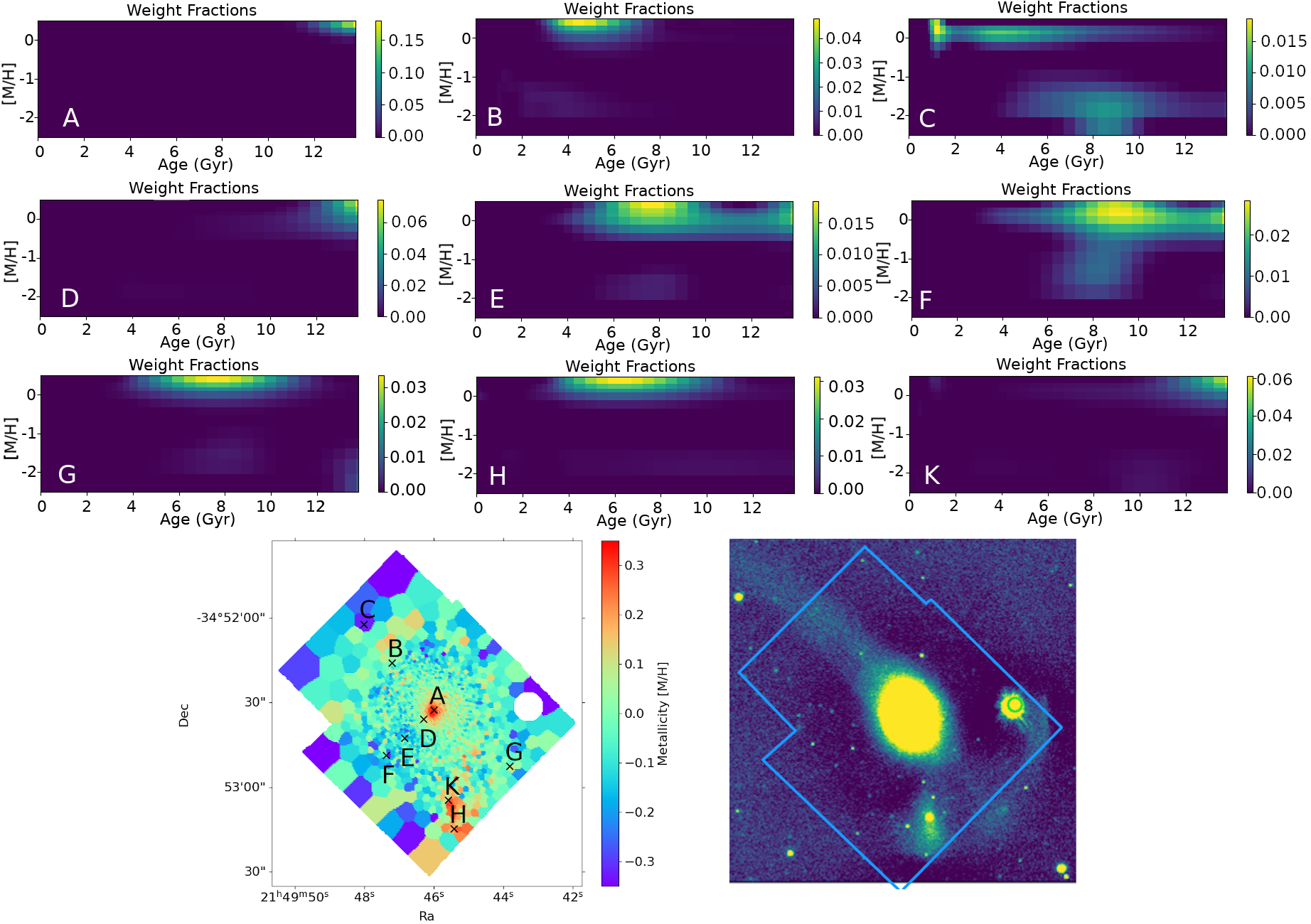}
    \caption{NGC\,7135 population sampling diagram. The upper nine panels display mass weighted metallicity space of NGC\,7135 for various regions. Corresponding regions are marked in the lower left panel with crosses marking the position extracted, and the corresponding letter. The lower right panel shows the same region as an unsharp masked image to highlight tidal features. Data for the unsharp masked image are taken from the VST ATLAS survey \protect{\citep{shanks2015vlt}}. The diagrams build a narrative in which a recent and ongoing merger creates large tidal features in NGC\,7135. There are also populations of far lower metallicity which are well mixed in the galaxy. These populations indicate historical mergers of high merger-mass ratio.}
    \label{var_regions}
\end{figure*}

We apply these principles to the population maps of NGC\,7135 in order to derive the history of formation and evolution. In Figure \ref{var_regions}, nine regions are marked with sequential letters corresponding to population maps, which are similarly sequentially lettered, with maps taken from the Voronoi bin below the labelled cross. Each position marks an area of interest or standard uniformity across the maps of Figure \ref{stellar_properties} with which we can build a picture of the assembly and current status of NGC\,7135. Region `A' marks the core of NGC\,7135. Regions `B' and `C' sample the tidal tail clearly seen in the unsharp mask image (lower right panel of Figure \ref{var_regions}), with increasing galactocentric radius. Regions `D', `E', and `F' also sample with increasing galactocentric radius, however they do so outside of any prominent tidal features. These are assumed to be a `control' sample which are chosen to represent the underlying galaxy, though show signs of probing accreted material. Regions `G' and `H' sample the tidal regions opposite the tail, with `H' particularly covering unstripped remnants of the infalling galaxy. Finally region `K' covers the core of the infalling galaxy.

Starting with region `A', we see a very high metallicity, very old population associated with the galaxy core. This is to be expected and is commonly seen in galaxy cores \citep[see e.g.][]{guerou2016}. There is little obvious evidence for accreted populations as expected, as shown by the old and high metallicity population, and lack of any clear population bimodality.

Moving along the main tidal tail in region `B' we see a much younger population at high metallicity. When comparing to regions not associated with tidal features but at similar radius such as `E' and `F', we see that the population of `B' is not comparable to `E' or `F'. This is largely due to a lack of older material that would be expected to be associated with the host galaxy. Plausibly this is the result of the vast majority of the stellar material originating in the infalling galaxy and comprising the tidal tail, and thus the populations visible are instead associated with this infalling object, rather than original populations of NGC\,7135. A small amount of material is also visible as a young and metal poor population. This can be attributed to ex-situ material that merged onto either NGC\,7135 or the infalling galaxy in the past prior to the current merger, and thus shows a separate population signature.

As we move further out along the tidal tail to region `C', many of the features become more prominent. For one thing, the high metallicity population associated with the stripped material from the infalling galaxy remains.  Furthermore, low metallicity ex-situ populations increase in the fraction of contributed mass (as seen as a distinctly separate low metallicity population). Care must be taken in comparison due to colour normalisation differences on the plot, however the maximum low metallicity ex-situ fraction increases from $\sim$0.5\% in `B' to $\sim$1.0\% in `C', with a higher sum total of ex-situ material. This increase is to be expected, as ex-situ material commonly increases in fraction with galactocentric radius \citep{LaBarbera12, Martin18, davison2020eagle}. It is unclear whether this ex-situ population is associated with NGC\,7135 or the infalling galaxy, however it could plausibly be from both, as models of hierarchical growth suggest both galaxies would have undergone historical minor mergers in all but the rarest cases \citep{fakhouri2010merger}. A burst of star formation is also seen in the final Gyr history. This is suggestive of a rapid star formation event, most likely triggered as a result of the galaxy interactions. Following this, no star formation is noticed in any bin. A shutdown of star formation after a major merger is discussed widely in literature \cite[see e.g.][]{bekki2001galaxy,barr2007formation,cortesi2013planetary,querejeta2015formation, Puglisi2021}.

Region `D' samples an inner region of NGC\,7135. It shows similar populations as in `A', however extends slightly to lower ages as expected following galaxy population age gradients. Little to no ex-situ material is clear. Moving further out in radius, we come to region `E'. This also shows the expected populations previously seen in `A' and `D'. This time however there is a more significant low metallicity ex-situ population, which as mentioned previously is expected as one reaches regions further from the galaxy centre according to galaxy simulations. Also prominent in region `E' is a population of intermediate age and high metallicity stars. As shown below in region `H', this is almost certainly associated with the infalling galaxy.

Region `F' samples at a slightly greater radius than `E', again with more prominent features, though in similar positions to `E'. We see an increase in the low metallicity ex-situ population radially along the tidal tail (`A', `B' and `C') and well as radially in areas not associated with tidal features (`D', `E' and `F').

The final regions sample the galaxy shell and associated infalling object. Region `G' examines an area of tidal shell seemingly also originating from the infalling galaxy. The region almost identically matches `H' which is placed to examine the outskirts of the infalling object, in regions that have yet to be stripped. The fact that these two populations are quite so similar suggests they are of the same origin, and that the tidal shells and tails are the result of scattered accreted material from the infalling galaxy.

Finally region `K' examines the core of the infalling galaxy at approximately 0.5 effective radii from the centre of NGC\,7135. It shows a highly metal rich and old population with the exact tendencies of a galaxy nucleus. It shows largely the same properties as the nucleus of NGC\,7135, though with marginally lower metallicity and a greater extent in age, suggesting a lower mass. 

The velocity dispersion of region `K' (seen in Fig \ref{stellar_properties}) is at a much lower average velocity dispersion than the host galaxy, again suggesting a lower mass of the merging galaxy compared to NGC\,7135. This is curious considering its high metallicity. One explanation would be that the in-falling galaxy is the remnant of a galaxy core stripped of its halo, which would explain both its relatively high brightness and high metallicity. This is also supported by the large amounts of seemingly ex-situ gas that are seen in Figure \ref{gas_properties}, where this gas would have formed the outer regions of the infalling galaxy as explained further in section \ref{gas_props}. 

The velocity dispersion (Fig \ref{stellar_properties}) increases significantly midway between the accreting galaxy core and the host galaxy core. This further lends weight to the idea that material is accreting onto the host galaxy, as the high velocity dispersion area indicates a region where accreted material begins encountering large amounts of in-situ material, and the difference in velocities becomes more evident, inflating the velocity dispersion, prior to mixing.

In summary, the population maps are indicative of three distinct galaxy populations, in which two significant merger events are present. The first is ongoing, with an intact core of a second galaxy currently in close proximity to NGC\,7135, with material being stripped off, accreted onto NGC\,7135, and creating large tidal features. These make up the high metallicity populations at intermediate ages. Yet another population is consistently present, as a low metallicity, intermediate to old aged population. As discussed previously, chemical enrichment and mass-metallicity relations mean this population is not associated with either galaxy. Therefore we attribute these stars to older historical mergers, now mixed loosely with the main populations. It is unclear which of these two present galaxies these populations accreted to, however as mentioned previously, the ex-situ population is likely present in both galaxies independently, and was captured by each prior to this ongoing merger.

\subsection{Accreted Populations}\label{acc_pop}
As seen in Figure \ref{var_regions}, many bins display a bimodality in population distribution (see e.g. panels `B', `C', `E', `F', `G', and `H'). Such a strong separation in populations suggests stellar material being obtained from more than a single source. Galaxies not associated with the main galaxy will evolve with a different metallicity due to the mass metallicity relation. As such, when the galaxies merge, there will be a distinct separation in the Age-Metallicity relation of each galaxy. The most obvious explanation for the bimodal populations seen in Figure \ref{var_regions} would be the merger of a less massive, lower metallicity galaxy to the host galaxy or onto the infalling galaxy, beginning $\sim$ 10\,Gyr ago. Furthermore, the fact that the bi-modality of populations is seen at almost all positions across the galaxy outside of the cores (panels `B', `C', `E', `F', `G', and `H') suggests that this material has been well mixed and is distributed throughout the galaxy, with the exception of the two galaxy cores (see panels `A', `D', and `K').

\begin{figure}
    \includegraphics[width=\linewidth]{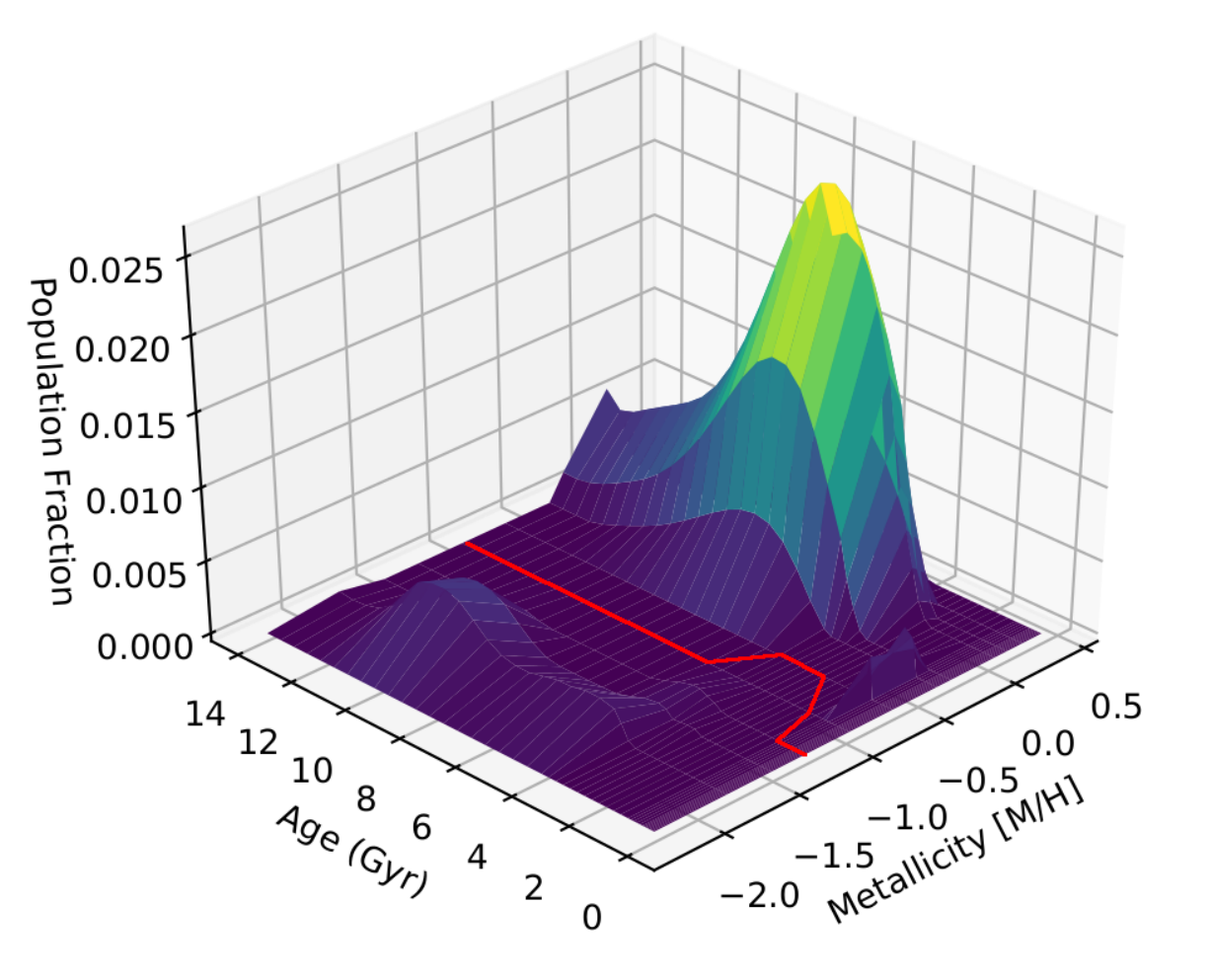}
    \caption{Population map of one bin projected on 3D axes. A line is sought for each map to bisect the lower metallicity population from the older using low saddle points. For this example, the path is marked by a red line.}
    \label{3dpath}
\end{figure}

To explore the population bi-modality, the fraction of stars not associated with the main host population was determined from each bin. To identify two discontinuous populations, a dividing line was sought across the population map, which would follow the lowest saddle points. This `path of least resistance' then divided the populations into two distinct sources; one being material from NGC\,7135 and the in-situ material of the infalling galaxy; and the other source being low metallicity populations accreted onto both galaxies at earlier times. This can be imagined as the valley between two hills, with the dividing line taking the natural path of a river at the lowest potential. This is visualised in Figure \ref{3dpath} with a red line showing the calculated separation path for one random bin, separating the populations into two sources.

\begin{figure*}
    \includegraphics[width=\linewidth]{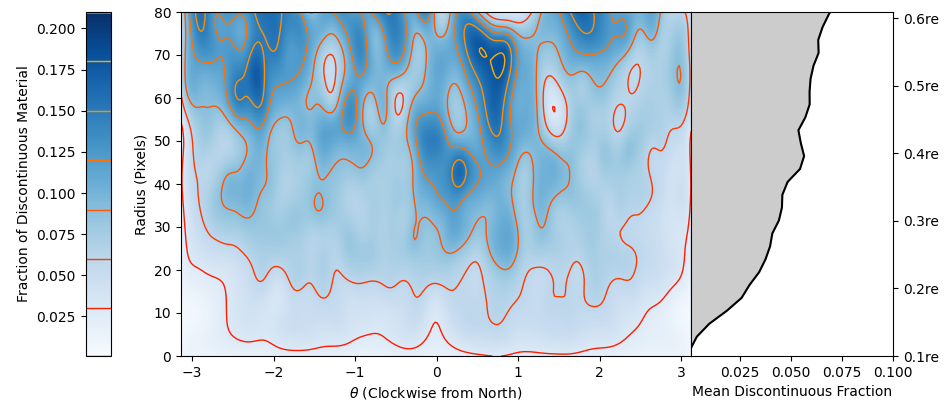}
    \caption{The left panel shows a polar oriented map of NGC\,7135. Blue colour shows the mass fraction of derived material not associated with the host galaxy population, with contouring shown in red-orange-yellow. The angle is shown with 0 radians as the North of the image and positive angle increase showing clockwise movement around the galaxy. Gaussian smoothing has been applied to show more clearly larger structures of ex-situ material. The radius from centre has been limited to include only radii in which a complete circle can be arranged within the image. The adjoining right-hand panel shows the same radial positions as the left side, however it shows the mean discontinuous mass fraction for a complete circle for the radii. Mean fraction was calculated using circular annuli of radius 3 pixels with a moving average. The effective radius is taken from table 1 of \protect\cite{marino2011nearby}. The fraction of accreted material increases with radius, with a roughly 7\% increase within 0.6 effective radii.}
    \label{bimodal}
\end{figure*}

Application of this to all bins provides a map such as in Figure \ref{bimodal}, where we can examine the fraction of stellar material associated with the lower metallicity source. Figure \ref{bimodal} shows a polar view of NGC\,7135 to better examine radial features. By examining fraction across the galaxy we can infer regions of higher or lower concentration of the accreted material. 

At the centre of NGC\,7135 we see no accreted material suggesting the core is dominated by in-situ stars. The density of accreted material rises with radius which is indicative of galaxy mergers depositing material on the outer regions of the galaxy. The material seems to be unevenly radially mixed, with proportionally higher quantities of ex-situ material deposited between 0 and 1 radians from North. This is likely a projection effect, as the area at the south of the galaxy (the left and right extents of Figure \ref{bimodal}) aligns with the previously mentioned high metallicity galaxy, with the stream of stellar material obscuring the host galaxy structure, and dominating the spectral light. 

\begin{figure*}
    \includegraphics[width=\linewidth]{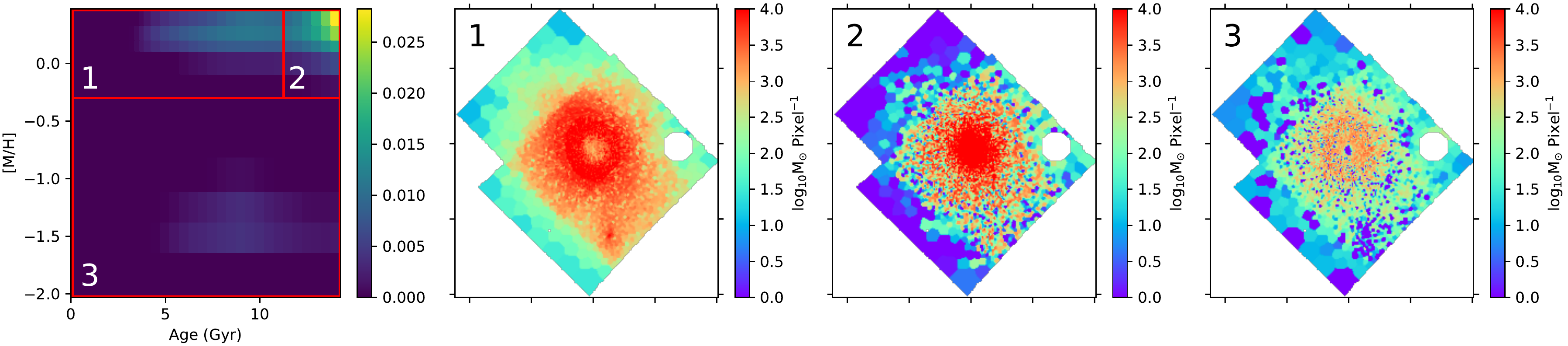}
    \caption{The first panel shows a general galaxy age-metallicity map. This is divided by the red boxes into 3 groups of populations to examine the mass associated with each area. Panel labels correspond to the numbers on the age-metallicity map. These show the divided nature of the populations, in which the intermediate age high metallicity population is more strongly associated with the infalling object and tidal features, whilst the older metal rich population is associated with the host galaxy.}
    \label{4pan_pops}
\end{figure*}

We can further see evidence of the division of the various populations by examining stellar mass estimates per population, determined with the division of the age-metallicity plane in combination with mass-to-light ratios. We show this in Figure \ref{4pan_pops}, with three regions of different populations separated roughly. Using mass to light ratios from \cite{thomas2003stellar}, we estimate the stellar mass per population division, per pixel. The panel labelled `1' corresponds to intermediate age stars with high metallicity which were associated with the infalling galaxy. This is confirmed in the first map in the Figure (panel 2) in which there is a noticeably higher stellar mass associated with the infalling object for only this population. This panel also encompasses much of the stellar material of NGC\,7135 near to the centre though at a slight distance, as is expected from standard galaxy age gradients. Though effects from the pointing overlaps are visible, it is notable that we see a small amount of material tracing the tidal tail and other tidally derived features. This suggests that the intermediate age material and tidal tail is associated with the infalling galaxy exclusively, though further data analysis from a higher resolution stellar model grid would be required for verification of this.

In the second labelled map (panel 3) we see that the most metal rich and oldest material is associated heavily with the host galaxy, with a strong gradient from the galaxy centre. This in-situ population is generally undisturbed and centrally concentrated, in comparison to the largely ex-situ population represented in the 1st map. Finally in the third labelled map (panel 4), we see again a gradient of stellar mass associated with the host galaxy. This third map shows only stars at far lower metallicities than the majority of the stellar material. This material is assumed to be low mass objects which have historically accreted to NGC\,7135, and are now well mixed into the galaxy. It should be noted that these are rigid divisions, and that the true population distributions from each object undoubtedly bleed over into the other divided regions (especially in regions `1' and `2').

\section{Discussion}
Analysis of the galaxy kinematics and gas of NGC\,7135 yielded evidence for both historical galaxy mergers, as well as an ongoing disruptive major merger. Despite the kinematics of past mergers being hidden (to the available resolution of data) due to mixing over time, ex-situ populations were extracted from the galaxy using full spectral fitting. This allowed for the identification of a well mixed low-metallicity stellar population relative to the larger fraction of higher metallicity stellar population. Considering expected enrichment patterns, this can only have occurred if either gas or stars (or both) originating in an ex-situ galaxy rapidly accreted or fully merged onto NGC\,7135. The lower metal content of this population made it distinct from the original population.

Potentially, all the stellar material in this population could have been created in-situ using gas that was accreted from another galaxy. This is highly unlikely however considering the specificity of the age and metallicity of the two distinct populations. Were these stars to be the product of new infalling gas, we would expect to see a mixing of the gas, and for the metallicity of new stars born after the merger event to be at a more intermediate metallicity. Instead, we see the two populations continuing to form stars without a sharp change in metallicity, thus the lower metallicity population stars are considered to be born ex-situ.

The bimodality of these stars allowed for clean separation of the ex-situ and in-situ populations. Thus the relative fraction of ex-situ material could be ascertained. This allowed for the exploration of ex-situ fraction with galactocentric radius, as shown in Figure \ref{bimodal}. The Figure shows a clear preference for ex-situ material to be located at the outer edges of the galaxy, with no detectable ex-situ material in the centre of the galaxy. This is akin to simulated results showing the same preference for ex-situ fraction increase with galactocentric radius \citep{schaye1014,Crain2015,rodrgom2016,davison2020eagle}, as well as observational studies showing the same principles \citep{forbes2011,Pillepich_2015,Oyarz_n_2019}. The mean ex-situ fraction measured for NGC\,7135 at approximately 0.6\,effective radii (the greatest extent captured by the MUSE image) is 7\%. This is only representative of the low metallicity populations from low-mass systems. Higher metallicity populations from mergers of smaller mass-ratio mergers would be disguised amongst in-situ populations.

Limitations of this technique largely arise from the ability to separate populations. At current resolutions of full spectral fitting techniques, populations must be wholly distinct in metallicity to be noticeably separable from the host population. Accreted material with age and metallicity similar to that of the host galaxy would be largely indistinguishable from the main population. Further limitations are the inability to directly distinguish between stars that are born ex-situ, and those born in-situ but of ex-situ material. As discussed above, these limitations are unlikely to be dominant in this scenario.

One interesting area to consider is the eventual fate of NGC7135. Will it retain some semblance of a spiral structure, or evolve into an S0 or elliptical galaxy? Conversion into an S0 galaxy seems to be a distinct possibility as S0 galaxies with coherent disk kinematics form through merger mechanisms, though the exact merger specifics continue to be debated within the community. Some evidence suggests that S0 galaxies are rarely expected to be formed through major mergers (<4:1 merger ratio) \citep{bournaud2005galaxy, lofthouse2016major}, with the conclusion given that major mergers are a plausible but non-dominant mechanism for early type formation. Conversely, other arguments suggest that S0 galaxies can indeed be formed from major mergers \citep{querejeta2015formation, querejeta2015formation2}. Furthermore major mergers can be shown to give rise to much of the inner structure often found in early types \citep{eliche2018formation}. Perhaps the most consistent agreement for the formation requirements of an S0 via mergers is the necessity for a misalignment of angular momentum between the in-situ and ex-situ accreted baryonic components \citep[see e.g.][]{sales2012origin}. Considering the existing baryonic misalignment present in NGC\,7135 in the form of a counter rotating disk, and considering the seemingly misaligned orbit of the ongoing merger, it is perhaps likely that the ongoing disruption will lead to NGC\,7135 tending towards S0 morphology. Plausibly the kinematics would increasingly reflect those of general spheroid galaxies as newly formed stars with an opposing angular momentum to the mean, and those recently accreted, would begin to reduce kinematic coherence. Though this is a distinct possibility, the true future of NGC\,7135 will remain unknown until more decisive techniques and modelling are developed. Due to the complex nature of the recent history of NGC\,7135, any predictions on future evolution are speculation.

\section{Conclusions}
We have used a Voronoi binned map of NGC\,7135 to explore kinematic stellar features such as velocity and velocity dispersion, as well as the distributions of stellar properties such as age and metallicity. Gas properties were also explored in regular bins, with both kinematic gas properties and gas distribution investigated. Gas was shown to be counter rotating compared to stellar material, with significant evidence of disturbance in the galaxy core. This along with population evidence shows a galaxy currently merging onto NGC\,7135. Despite gas being present, little to no current star formation was identified. ALMA data of the galaxy core points to a star formation rate of only $0.025M_{\sun}\,\mathrm{yr}^{-1}$ assuming normal depletion times. Strong LINER emission likely obscures emission associated with star formation and as such a higher SFR cannot be ruled out.

During population analysis of NGC\,7135 from data provided by the SOSIMPLE project, we have identified both historic and ongoing merger activity. This was achieved using a `full spectral fitting' method to disentangle strong bi-modalities in stellar populations. We show that in a snapshot of a `single' galaxy, we are in reality witnessing the product of three distinct galaxy populations.

An ongoing merger or large accretion event is clear from the stellar kinematic maps, showing a distinct area of stellar material not associated with the host galaxy, but interacting with the galaxy structure. Likewise in gas maps we see large velocity dispersion in areas where ex-situ infalling gas interacts with in-situ gas. 

At least one historical large merger event took place at 6-10\,Gyr ago according to star-formation history derived by full spectral fitting. This potentially provided gas with lower enrichment with which NGC\,7135 birthed stars of lower metallicity; however the timeline of stellar ages, matched with the likely merger date makes it highly likely that most, if not all of the stars belonging to this population are ex-situ stars originating in another galaxy. Considering there is no discernible change in the host population metallicity of new stars born after the merger, we assume that all lower metallicity population stars are ex-situ in origin. The timeline of star formation history suggests that this merger caused a general shut-down of star formation in NGC\,7135, not long after the merger event.

We calculate the fraction of the ex-situ material as a function of galactocentric radius, finding a steep increase in ex-situ material as we probe further to the outskirts of the galaxy. The centre of the galaxy exhibits no signs of ex-situ material, whilst by 0.6 effective radii, this fraction is at 7\%. This is in common with literature expectations of `two phase' galaxy assembly seen both observationally and in simulation, where ex-situ material is preferentially deposited on the outskirts of a galaxy.

Many more SOSIMPLE galaxies are available from the survey, with much left to explore.

\section{Acknowledgements}
Many thanks to an anonymous referee for useful comments. This work was completed with the support of the ESO studentship program and the Moses Holden Scholarship. BH acknowledges support by the DFG grant GE625/17-1 and DLR grant 50OR1911. Based on observations collected at the European Southern Observatory under ESO programme 0103.A-0637(A).
\section{Data Availability}
The data described in this article are accessible via the ESO archive of MUSE data.
\bibliographystyle{mnras}
\bibliography{biblio}

\bsp	
\label{lastpage}
\end{document}